 \documentclass[12pt, draftclsnofoot, onecolumn]{IEEEtran}  
 
\usepackage{amsmath,amssymb,graphicx,epstopdf,cite,subfigure,epsf}
\usepackage{psfrag}
\usepackage{amssymb}
\usepackage{amsmath}
\usepackage{bbm}
\usepackage{dsfont}
\usepackage{pifont}
\usepackage{cite}
\usepackage{graphics}
\usepackage{epsfig}
\usepackage{amscd}
\usepackage{accents}
\usepackage{multirow}
\usepackage{epstopdf}
\usepackage{algorithm,algpseudocode}
\usepackage{color}
\usepackage{varwidth}

\newlength{\dhatheight}

\title{ Efficient Beam Training and Channel Estimation  for Millimeter Wave Communications Under Mobility  }

\author{\IEEEauthorblockN{Sun Hong Lim, Jisu Bae, Sunwoo Kim, Byonghyo Shim*, and Jun Won Choi }\\
\IEEEauthorblockA{Hanyang University, *Seoul National University \\
Email: \ shlim, jsbae@spo.hanyang.ac.kr, remero@hanyang.ac.kr, *bshim@snu.ac.kr, junwchoi@hanyang.ac.kr}

}

\begin{document}

\maketitle

\begin{abstract}
In this paper, we propose an efficient beam training technique for millimeter-wave (mmWave) communications. When some mobile users are under high mobility, the beam training should be performed  frequently to ensure the accurate acquisition of the channel state information. In order to reduce the resource overhead caused by frequent beam training, we introduce a dedicated beam training strategy which sends the training beams separately  to  a specific high mobility user (called a target user) without changing the periodicity of the conventional beam training. The dedicated beam training requires small amount of resources  since the training beams can be optimized for the target user. In order to satisfy the performance requirement with low training overhead, we propose the optimal training beam selection strategy which finds the best beamforming vectors yielding the lowest channel estimation error based on the target user's probabilistic channel information. Such dedicated beam training is combined with the greedy channel estimation algorithm that accounts for sparse characteristics and temporal dynamics of the target user's channel. Our numerical evaluation demonstrates that the proposed scheme can maintain good channel estimation performance with significantly less training overhead compared to the conventional beam training protocols.
 \end{abstract}

\begin{IEEEkeywords}
Millimeter wave communications, Beam training, Beam tracking, Mobility, Channel Estimation
\end{IEEEkeywords}

\IEEEpeerreviewmaketitle

\section{Introduction}
In recent years, wireless communications using millimeter-wave (mmWave) frequency band has received great deal of attention as a means to meet ever-increasing throughput demand of next generation communication systems  \cite{rappaport,niu,overview}. Basically, millimeter-wave band covers the frequency band ranging from 30 GHz to 300 GHz, which is much higher than the frequency band in current cellular systems.
There is a vast amount of bandwidth that has not been explored by the current communication systems and thus  mmWave communications can be a promising solution to the ever-increasing throughput requirements.
Notwithstanding the great promise and potential benefit, there are some drawbacks and obstacles that need to be addressed for the commercialization of mmWave-based communication systems. One major obstacle is the significant path loss of mmWave channels \cite{chan60,chan602,chan_rappa}. Compared to the conventional communication systems using microwave radio waves, mmWave band experiences high atmospheric attenuation when the transmit signal is absorbed by gas and humidity. Additionally, there would be a significant path loss when the signal is blocked by obstacles such as building, foliage, and user's body \cite{niu}.
The key enabler to overcome this drawback is the \emph{beamforming} in which two communication entities transmit and receive the signals with appropriately adjusted phase and amplitude using an array of antennas \cite{overview,roh,beamforming,chan_est}. Since the wavelength in the mmWave communication systems is in the range of one to ten millimeters, a large number of antenna elements can be integrated into a small form factor, making  the highly directional beamforming to compensate for the large path loss of mmWave channels possible. 

Over the years, various beamforming strategies, such as hybrid beamforming, switched beamforming, and multi-stage beamforming, have been introduced  \cite{hybrid,donno,analog,wang}.
In a nutshell, conventional beamforming protocol consists of two major steps. The {first step} is the \emph{beam training}  \cite{overview,chan_est}. In this step, the base-station  transmits the known training symbols to the mobile users periodically. These training symbols are transmitted at some designated directions using the {\it training beams} (this beams are often called the {\it sounding beams} \cite{sounding}).    Using the received beams, a user acquires the channel state information (CSI) which corresponds to the channel gains for all pairs of the transmit antenna and receive antenna. In the widely used \emph{beam-cycling} scheme, for example, the base-station sequentially transmits the $N$ beams steered at the equally distributed directions. In the mobile terminal, each user estimates its own CSI and then feeds back a certain function of its own CSI (e.g., a precoding index selected from the codebook) to the base-station.
Then, the base-station performs the  precoding  of data symbols, which corresponds to the second step of beamforming. In this step, the base-station transmits the data symbols using the precoding matrix  designed to maximize the system throughput \cite{precoding, sparse_precoding}.

While high directional beamforming is an effective means to improve the system performance, the system becomes inefficient when the users' locations change in time or the position and heading angle of the devices vary relative to the base-station.
This is because the beam training of the mobile users should be performed more frequently to track the CSIs of the mobile users. Since the beams are shared by all co-scheduled users, even when only a fraction of users are under mobility, beam training should be performed frequently, imposing substantial overhead in resource utilization. 
Recently, various attempts have been made to enhance the beam training efficiency. These include the adaptive beamtraining \cite{chan_est}, codebook-based beam switching \cite{beamtraining}, simplex optimization-based beam training \cite{yuan}, beamcoding approach \cite{tsang}, and multi-level beam training \cite{wang}. 
When  users under high mobility exist in the cell, the training beams should be transmitted during data transmission period in order to track their time-varying channels. In particular, the beam transmission strategy for handling such high mobility users is often called {\it beam tracking} \cite{heathtr,palacios,sounding}. 
For example, in \cite{heathtr} and \cite{palacios}, the known pilot symbol is periodically transmitted using the single training beam used to precode the data symbols. However, use of a single training beam does not promise good tracking performance.  In \cite{sounding}, the directional multiple training beams are  designed accounting for the temporal statistics of angle of departure (AoD) and angle of arrival (AoA). In \cite{sung},  optimal beam training protocol has been derived from the partially observable Markov decision process (POMDP) framework.  Various channel tracking algorithms have also been proposed.  Kalman filters of various structures were applied to jointly estimate temporally correlated AoD, AoA, and the channel gain \cite{heathtr, mmwave_tracking1,mmwave_tracking2,he}.  While Kalman filters are effective in tracking time-varying channels, they suffer from performance degradation when the number of paths  in channel is large. It has been shown in \cite{chan_est} and \cite{beamspace} that the compressed sensing (CS) recovery algorithms can effectively estimate the AoDs and AoAs for multi-path channels \cite{cs_magazine,cs_shim}. The CS-based channel estimation has been extended to track the time-varying mmWave channel exploiting the temporal channel correlation \cite{sounding,tera,madhow}.  In \cite{sounding}, approximate message passing (AMP) algorithm has been employed for channel tracking.  In \cite{tera}, probabilistic knowledge on the AoD and AoA obtained from the channel estimation in the previous step was incorporated as a priori for the channel estimation in the current step. In \cite{madhow}, multi-path channel parameters are estimated iteratively based on maximum likelihood criterion.

The aim of this paper is to propose a new beam training strategy to support the mobility scenario in mmWave communications. 
Inspired by the user-specific pilot transmission scheme in the 4G long-term evolution (LTE) standard \cite{3gpp211}, we consider the \textit{dedicated beam training} designed to  facilitate the beam training only for the high mobility user without changing the periodicity of the conventional beam training. 
While all users in a cell are supported by the conventional beam referred to as the {\it common beam}, the proposed {dedicated beam} takes a complementary role of supporting the users under high mobility which cannot acquire the CSI accurately only using the common beam training.
Since the dedicated beams are intended only for the high mobility users (in the sequel we called it {\it target user}), it is possible to quickly adapt the direction and periodicity of the training beams to the channel of the target user and thus improve the radio resource utilization for training. 

This paper mainly focuses on problems of designing the training beams and channel estimation for the dedicated beam training.
First, we present an optimal training beam search algorithm to choose the best $N_d$ beam indices minimizing the channel estimation error from the beam codebook. Our beam design steers the $N_d$ training beams adaptively towards the target user at the directions accounting for the probabilistic knowledge of the target user's channel delivered by our channel estimation. Our evaluation shows that  at least more than two dedicated training beams are needed to ensure good channel estimation performance and the dual beam transmission   requiring only two channel uses  ($N_d=2$) can achieve the performance comparable to using $N_c = 32$ common beams with the same periodicity. We also show that for dual beam transmission, our beam selection algorithm can be implemented at low complexity through one dimensional grid search.   Second, we propose a new sequential CS-based channel estimation algorithm for the dedicated beam training. The proposed method builds upon the greedy sparse recovery algorithm,  (i.e.,  the orthogonal matching pursuit (OMP)  \cite{omp})  which successively finds the pair of AoD and AoA until it finds the pairs corresponding to all channel paths. Using the statistical model describing temporal correlation of mmWave channel under mobility, we derive the enhanced greedy channel estimation algorithm that exploits the temporal structure of channel.  Note that proposed channel estimator incorporates the priori distribution of AoD and AoA in finding the pair of AoD and AoA and produces their probabilistic distribution as an output of our channel estimator. This output is used to perform the beam selection and the channel estimation for the next beam training period.
%
  We demonstrate from numerical experiments that the combination of our channel estimation algorithm and the dedicated beam training yields the performance comparable to the conventional beam cycling with much smaller training overhead.  
  
Our work is different from the existing beam tracking methods \cite{heathtr,palacios,sounding,mmwave_tracking1,mmwave_tracking2,sung,he} in that based on the probabilistic knowledge on the AoD and AoA (reported by the channel estimator), we find the multiple best  training beams minimizing the  channel estimation error directly. Furthermore, we introduce the new sequential greedy algorithm that leverages both the temporal correlation of AoD and AoA and the channel sparsity. Note that such type of greedy channel estimator has not been proposed in the literature yet, to our best knowledge.

The rest of this paper is organized as follows; 
In section \ref{sec:sys_model}, we introduce the system and channel models for mmWave communications. In section \ref{sec:proposed_scheme}, we describe the proposed beam tracking method. In section IV, we present  the simulation results, and conclude the paper in section V.

The notations to be used in the rest of paper are as follows. 
Operations $(\cdot)^{T}$, $(\cdot)^{H}$, $(\cdot)^{*}$, and $(\cdot)^{-1}$ denote transpose, Hermitian, conjugate, and inverse operations, respectively.  $\mathbf{A}\otimes \mathbf{B}$ is Kronecker product of two matrices $\mathbf{A}$ and $\mathbf{B}$. $E[X]$ and $Var(X)$ denotes the expectation and variance of the random variable $X$.
%
$(\mathbf{X})_{i,j}$ denotes the $(i,j)$th element of the matrix $\mathbf{X}$ and $(\mathbf{x})_{i}$ is the $i$th element of the vector $\mathbf{x}$.
$\mathbf{X}_{\Omega}$ is the submatrix of $\mathbf{X}$ that contains the columns as specified in the set $\Omega$.
and $(\mathbf{x})_{\Omega}$ is the vector constructed by picking the elements from the vector $\mathbf{x}$ as specified in the set $\Omega$.
$A \succeq B $ implies that $(A-B)$ is positive semidefinite. ${\rm vec}(\mathbf{X})$ is the vectorization operation of the matrix $\mathbf{X}$. $\mathcal{CN}(\mu, \Sigma)$ denotes the complex multivariate Gaussian distribution with the mean $\mu$ and covariance matrix $\Sigma$.

\section{MmWave Channel Model and Conventional Beam Training} \label{sec:sys_model}

%

In this section, we briefly describe the system model for mmWave communications.
We first present the mmWave channel model and then discuss the basic procedure of the conventional beam training and channel estimation.

\begin{figure} [t]
 \centering
 \includegraphics[width=5in]{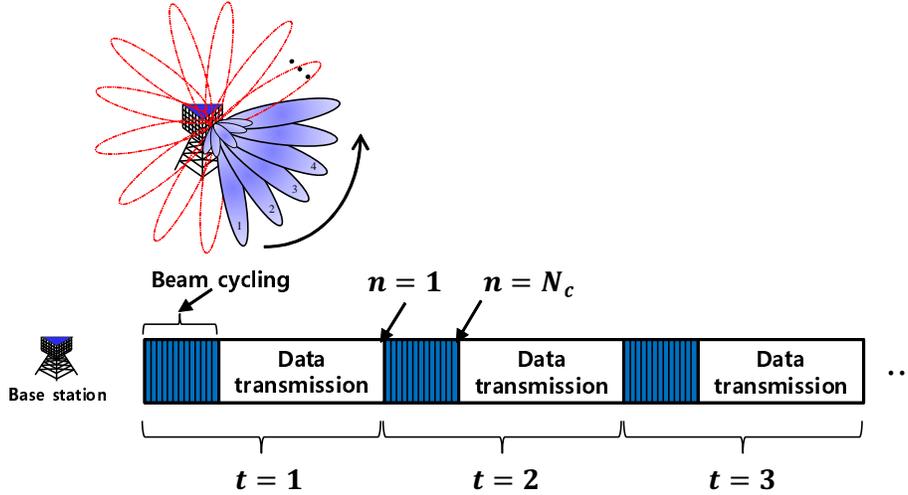}
 \caption{Frame structure of the conventional common beam training.} \label{fig:common}
\end{figure}

\subsection{System Description}

Consider the mmWave downlink where a base-station equipped with $N_b$ antennas is serving $P$ users with $N_m$ antennas. 
The frame structure for the conventional common beam training is depicted in Fig. ~\ref{fig:common}.
In each beam training period, the base-station transmits $N_c$ beams  sequentially at the designated directions. Each user estimates its own channel based on the received training signals and calculates the precoding matrix (or picks the precoding index from the codebook) using the channel information.  Then, each user feeds back the precoding matrix (or selected precoding index) to the base-station.
Once the beam training procedure finishes, the base-station transmits the precoded data symbols to the users during the data transmission period. Note that the beam training procedure is performed periodically to keep track of channels for all users. 


\subsection{MmWave Channel Model} \label{subsec:channel_model}
In general, the mmWave channel  from the base-station to the $p$th user  can be expressed as the $N_m \times N_b$ channel matrix
$\mathbf{H}_{p,t,n}$ where the subscripts $t$ and $n$ represent the $t$th beam training period and the $n$th beam transmission, respectively
(see Fig. ~\ref{fig:common}). The $(i,j)$th element of $\mathbf{H}_{p,t,n}$  represents the channel gain from the $j$th antenna of the base-station to the $i$th antenna of the user.  The angular domain representation of the channel matrix $\mathbf{H}_{p,t,n}$ is expressed as  \cite{chan_est,beamspace}
\begin{align}
  \mathbf{H}_{p,t,n} & = 
  \sum\limits_{l=1}^L\alpha_{p,t,n,l}\mathbf{a}^{(m)}(\theta^{(m)}_{p,t,n,l})\left(\mathbf{a}^{(b)}(\theta^{(b)}_{p,t,n,l})\right)^{H} \\
  & = \mathbf{A}_{p,t,n}^{(m)} \mathbf{\Lambda}_{p,t,n} (\mathbf{A}_{p,t,n}^{(b)})^{H}, \label{eq:physical_channel}
\end{align}
where
\begin{align}
 \mathbf{A}_{p,t,n}^{(m)}  &=  \begin{bmatrix}\mathbf{a}^{(m)}(\theta^{(m)}_{p,t,n,1}) & \cdots & \mathbf{a}^{(m)}(\theta^{(m)}_{p,t,n,L}) \end{bmatrix} \\
  \mathbf{\Lambda}_{p,t,n}  &= \begin{bmatrix}\alpha_{p,t,n,1}& 0 & 0 \\ 0 & \ddots & 0 \\ 0 & 0 & \alpha_{p,t,n,L} \end{bmatrix} \\
  \mathbf{A}_{p,t,n}^{(b)} &= \begin{bmatrix}\mathbf{a}^{(b)}(\theta^{(b)}_{p,t,n,1}) & \cdots & \mathbf{a}^{(b)}(\theta^{(b)}_{p,t,n,L}) \end{bmatrix}.
\end{align}
 Note that $L$ is the total number of the paths,  $\alpha_{p,t,n,l}$ is the channel gain for the $l$th path, and $\theta^{(b)}_{p,t,n,l}$ and $\theta^{(m)}_{p,t,n,l}$ are the AoD and AoA for the $l$th path where
\begin{align}
  \theta^{(b)}_{p,t,n,l} & = \sin(\phi_{p,t,n,l}^{(b)}), \nonumber \\
  \theta^{(m)}_{p,t,n,l} & = \sin(\phi_{p,t,n,l}^{(m)}) \nonumber
\end{align}
where $\phi_{p,t,n,l}^{(b)}$, $\phi_{p,t,n,l}^{(m)}\in[-\frac{\pi}{2},\frac{\pi}{2}]$ are the  angles in radian for AoD and AoA, respectively. Notice that $ \mathbf{a}^{(b)}(\theta) $ and $\mathbf{a}^{(m)}(\theta)$   are the steering vectors  for the base-station and user, respectively. That is,
\begin{align}
 \mathbf{a}^{(b)}(\theta) & = \frac{1}{\sqrt{N_{b}}}\left[1,e^{\frac{j2\pi d_b \theta}{\lambda}},e^{\frac{j2\pi2d_b\theta}{\lambda}},\cdots,e^{\frac{j2\pi (N_{b}-1)d_b \theta}{\lambda}}\right]^{T} \nonumber \\
  \mathbf{a}^{(m)}(\theta) & = \frac{1}{\sqrt{N_{m}}}\left[1,e^{\frac{j2\pi d_m \theta}{\lambda}},e^{\frac{j2\pi2d_m\theta}{\lambda}},\cdots,e^{\frac{j2\pi (N_{m}-1)d_m\theta}{\lambda}}\right]^{T}, \nonumber
\end{align}
where $d_b$ and $d_m$ are the distances between the adjacent antennas for the base-station and  user, respectively, and $\lambda$ is the signal wavelength. 
Although the AoD and AoA have real values in the range $[-1,1]$, they can be approximated to the discrete values by quantizing them on the uniform grid of $M_b$ and $M_m$ bins.  That is, 
\begin{align}
\theta^{(b)}_{p,t,n,l} \in& \left[-1,-1+2\frac{1}{M_b},\ldots,-1+2\frac{(M_b-1)}{M_b}\right] \nonumber \\
\theta^{(m)}_{p,t,n,l} \in& \left[-1,-1+2\frac{1}{M_m},\ldots,-1+2\frac{(M_b-1)}{M_m}\right].
\label{eq:angular_model}
\end{align}
Adopting the quantized channel representation referred to as virtual channel representation, we have \cite{chan_est,beamspace,tera}
\begin{align} \label{eq:canon}
  \mathbf{H}_{p,t,n} & = \mathbf{A}^{(m)}\mathbf{H}_{p,t,n}^{(v)} \left(\mathbf{A}^{(b)}\right)^{H},
\end{align}
where
\begin{align}
\mathbf{A}^{(b)} 
&=\Big[ \mathbf{a}^{(b)}(-1),\mathbf{a}^{(b)}\Big(-1+\frac{2}{M_b}\Big),\ldots,\mathbf{a}^{(b)}\Big(-1+2\frac{M_b-1}{M_b}\Big)\Big] \nonumber \\
\mathbf{A}^{(m)} 
&=\Big[ \mathbf{a}^{(m)}(-1),\mathbf{a}^{(m)}\Big(-1+\frac{2}{M_m}\Big),\ldots, \mathbf{a}^{(m)}\Big(-1+2\frac{M_m-1}{M_m}\Big)\Big]. \nonumber
\end{align}
The $(i,j)$th element of  $\mathbf{H}_{p,t,n}^{(v)}$ is the channel gain corresponding to the $i$th angular bin for the AoA and the $j$th angular bin for the AoD. If there exist $L$ multi-paths, only $L$ entries of $\mathbf{H}_{p,t,n}^{(v)}$ have dominant values and the rest entries are close to  zero. Since the number of multi-paths $L$ is in general much smaller than that of elements in $\mathbf{H}_{p,t,n}^{(v)}$,
the channel matrix $\mathbf{H}_{p,t,n}^{(v)}$ can be readily modeled as a sparse matrix in the angular domain.

\subsection{Conventional Beam Training and Channel Estimation} \label{subsec:BT_and_channel_est}

As mentioned, during the beam training period, the base-station transmits the known symbols over the $N_c$ consecutive time steps.
In the $n$th beam transmission, for example, the base-station transmits the known symbol  $s_n$ through the beamforming vector $\mathbf{f}_{t,n}$.
The corresponding received vector for the $p$th user is given by
\begin{align}
	\mathbf{r}_{p,t,n} = \mathbf{H}_{p,t,n} \mathbf{f}_{t,n} s_{n}  + \mathbf{n}_{p,t,n},
\end{align}
where $\mathbf{f}_{t,n}$ is the $N_b\times1$ beamforming vector and $\mathbf{n}_{p,t,n}$ is the $N_m \times 1$ Gaussian noise vector $\mathcal{CN}(0,\sigma_n^2\mathbf{I})$.
Since $s_n$ is the known symbol, we let $s_{n}=1$ in the sequel for simplicity.
When generating the beamforming vector, the {\it beam cycling} scheme in which the base-station transmits $N$ training beams at equally spaced directions is popularly used \cite{overview} (see Fig.\ref{fig:common})
\begin{align}
\mathbf{f}_{t,1} &= \mathbf{a}^{(b)}(-1) \\
\mathbf{f}_{t,2} &= \mathbf{a}^{(b)}\left(-1+2\frac{1}{N_c}\right) \\
\vdots \nonumber \\
\mathbf{f}_{t,N_c} & = \mathbf{a}^{(b)}\left(-1+2\frac{N_c-1}{N_c}\right).
\end{align}
Note that such beam transmission can be implemented by the analog beamformer  with a single RF chain.  
Clearly, the larger the number of beams $N_c$, the better the quality of the channel estimation would be, but larger $N_c$ leads to higher resource overhead.
%
%
The $p$th user  multiplies  the $N_m \times 1$  combining vector to $ \mathbf{r}_{p,t,n}$ by $N_m$ times, i.e.,
\begin{align}
\mathbf{y}_{p,t,n} &=  \mathbf{W}_{p,t}^{H} \mathbf{r}_{p,t,n} \\
& = \mathbf{W}_{p,t}^{H} \mathbf{H}_{p,t,n} \mathbf{f}_{t,n} s_{n} + \mathbf{W}_{p,t}^{H} \mathbf{n}_{p,t,n},
\end{align}
where $\mathbf{W}_{p,t} = [\mathbf{w}_{p,t,1}, ..., \mathbf{w}_{p,t,N_m}]$ is the combining matrix. 
For example, we can use $N_m \times N_m$ DFT matrix for $\mathbf{W}_{p,t}$.  

Next, we describe the basic channel estimation algorithm used for the conventional beam training. 
Using the vectors $\mathbf{y}_{p,t,1},...,\mathbf{y}_{p,t,N_c}$ collected from the $t$th beam training period,
the $p$th user  estimates the channel matrix $\mathbf{H}_{p,t,n}$.
We assume that the channel matrix $\mathbf{H}_{p,t,n}$ is fixed during the single beam training period, i.e.,
 $\mathbf{H}_{p,t} = \mathbf{H}_{p,t,i}$  $(i=1,2,,...,N_c)$.
Let $\mathbf{Y}_{p,t} = [\mathbf{y}_{p,t,1},...,\mathbf{y}_{p,t,N_c}]$, then
\begin{align}
 \mathbf{Y}_{p,t}&
=  \mathbf{W}_{p,t}^{H}  \mathbf{H}_{p,t} \mathbf{F}_t +  \mathbf{W}_{p,t}^{H}  \mathbf{N}_{p,t} \\
& =   \mathbf{W}_{p,t}^{H}  \mathbf{H}_{p,t} \mathbf{F}_t  + \mathbf{N}'_{p,t},
 \label{eq:rpt}
 \end{align}
 where $\mathbf{F}_t = \begin{bmatrix} \mathbf{f}_{t,1} & \cdots & \mathbf{f}_{t,N_c}  \end{bmatrix}$ and  $\mathbf{N}'_{p,t} = \mathbf{W}_{p,t}^{H}  \mathbf{N}_{p,t}$.
Since the channel matrix $ \mathbf{H}_{p,t}$ can be represented by a small number of parameters in  the angular domain, we use the angular channel representation in (\ref{eq:physical_channel}), and as a result
%
%
\begin{align} \label{eq:yv}
\mathbf{Y}_{p,t} &= \mathbf{W}_{p,t}^{H} \mathbf{A}_{p,t}^{(m)} \mathbf{\Lambda}_{p,t} (\mathbf{A}_{p,t}^{(b)})^{H}  \mathbf{F}_t +    \mathbf{N}'_{p,t} \\
 & = \sum\limits_{l=1}^L\alpha_{p,t,l}\mathbf{W}_{p,t}^{H} \mathbf{a}^{(m)}(\theta^{(m)}_{p,t,l})\left(\mathbf{a}^{(b)}(\theta^{(b)}_{p,t,l})\right)^{H} \mathbf{F}_t  +    \mathbf{N}'_{p,t} \label{eq:an}
 \end{align}
Note that the subscript $n$ is dropped from the channel parameters  $\theta^{(b)}_{p,t,l}$, $\theta^{(m)}_{p,t,l}$, and $\alpha_{p,t,l}$ since we assume that the CSI is invariant during the beam training period.
%
Due to the nonlinearity of AoD and AoA with respect to the received vector, joint  estimation of the parameters $\alpha_{p,t,1}, ..., \alpha_{p,t,L}$, $\theta^{(b)}_{p,t,1}, ..., \theta^{(b)}_{p,t,L}$, and  $\theta^{(m)}_{p,t,1}, ..., \theta^{(m)}_{p,t,L}$ tends to be computationally infeasible.  Alternatively, one can use the virtual channel representation, $ \mathbf{H}_{p,t}=  \mathbf{A}^{(m)}\mathbf{H}_{p,t,n}^{(v)} \left(\mathbf{A}^{(b)}\right)^{H}$ in (\ref{eq:rpt}) and construct  the observation vector $\mathbf{y}_{p,t}$  by stacking the columns of $\mathbf{Y}_{p,t}$ as   \cite{chan_est}
\begin{align}
\mathbf{y}_{p,t}  =& {\rm vec}(\mathbf{Y}_{p,t}) \nonumber \\
=&  {\rm vec}( \mathbf{W}_{p,t}^{H} \mathbf{A}^{(m)}\mathbf{H}_{p,t}^{(v)} \left(\mathbf{A}^{(b)}\right)^{H} \mathbf{F}_t +  \mathbf{N}'_{p,t}) \nonumber \\
=& \left( \mathbf{F}_t^{T} \left(\mathbf{A}^{(b)}\right)^{*} \right)\otimes \left( \mathbf{W}_{p,t}^{H} \mathbf{A}^{(m)}\right) {\rm vec} \left( \mathbf{H}_{p,t}^{(v)}\right) + {\rm vec}(\mathbf{N}'_{p,t}) \nonumber \\
=& \mathbf{\Phi}_{p,t} \mathbf{h}_{p,t} + \mathbf{n}_{p,t}, \label{eq:feq}
\end{align}
where
$\mathbf{\Phi}_{p,t} = \left( \mathbf{F}_t^{T} \left(\mathbf{A}^{(b)}\right)^{*} \right)\otimes \left( \mathbf{W}_{p,t}^{H} \mathbf{A}^{(m)}\right)$,
$\mathbf{h}_{p,t} = {\rm vec} \left( \mathbf{H}_{p,t}^{(v)}\right) $, and
$\mathbf{n}_{p,t} = {\rm vec}\left(\mathbf{N}'_{p,t}\right)$.
%
Note that the channel estimation is equivalent to the estimation of the unknown vector $\mathbf{h}_{p,t}$ from the received vector $\mathbf{y}_{p,t}$ in (\ref{eq:feq}). According to the definition $\mathbf{h}_{p,t} = {\rm vec} \left( \mathbf{H}_{p,t}^{(v)}\right) $, the index of elements in $\mathbf{h}_{p,t}$ represents the  pair of AoD and AoA and its value represents the channel gain. 
In practical scenarios, we cannot use  many training beams due to the limitation of resources, which in turn means that the number of rows in $\mathbf{\Phi}_{p,t}$ should be less than the number of columns. While obtaining an accurate estimate of $\mathbf{h}_{p,t}$ in an underdetermined systems is in general very difficult, 
we can accurately recover $\mathbf{h}_{p,t}$ by exploiting the sparsity of the channel vector $\mathbf{h}_{p,t}$.
In estimating the channel $\mathbf{h}_{p,t}$, we can basically use any sparse recovery algorithm \cite{cs_magazine,cs_shim} including OMP \cite{omp}. (see Algorithm \ref{alg:omp}.)

\begin{algorithm} [t]
	\caption{OMP algorithm to estimate mmWave channel}
	\textbf{Initialize : }$\mathbf{r}^{(0)} = \mathbf{y}_{p,t}, \Omega_0 = \{\}$\\
	\textbf{Output : } Support set $\Omega$ and channel gain $\hat{\mathbf{h}}_{\Omega}$
	\begin{algorithmic}[1]
		\For {$i=1$ to $L$$...$}
        \State \begin{varwidth}[t]{\linewidth} Select the index maximizing the magnitude of the inner product between $\psi_i$ \par and  $\mathbf{r}^{(i-1)}$ : 
        \end{varwidth}
         \begin{align*}
		{i^*} = \arg\underset{i}{\max} ~  | \psi_i^H \mathbf{r}^{(i-1)} |^2,
		\end{align*}
        \hspace{0.5cm} where $\psi_i$ is the $i$th column of $\mathbf{\Phi}_{p,t}$.     
       		\State Update the support set $\Omega$:
		\begin{align*}
		&\Omega_{i} = \Omega_{i-1} \cup {i}^*
		\end{align*}
		\State Update the signal amplitude $\hat{h}_{\Omega}$ and the residual $r^{(i)}$;
		\begin{align*}
		&\hat{\mathbf{h}}_{\Omega_i} = \left( (\mathbf{\Phi}_{p,t})_{\Omega_i}^H (\mathbf{\Phi}_{p,t})_{\Omega_i}  \right)^{-1} (\mathbf{\Phi}_{p,t})_{\Omega_i}^H \mathbf{y}_{p,t}
		\\ &\mathbf{r}^{(i)} =  \mathbf{y}_{p,t} - (\mathbf{\Phi}_{p,t})_{\Omega_i} \hat{\mathbf{h}}_{\Omega_i}
		\end{align*}
		\EndFor
 \end{algorithmic}
 \label{alg:omp}
\end{algorithm}

\section{Proposed Dedicated Beam Training For Mobility Scenario} \label{sec:proposed_scheme}

In this section, we present the new beam training technique to support the users  under mobility. 
Since the common beam training should serve all users 
in a cell, a large number of training beams to cover wide range of direction are needed.
As mentioned, when there are moving users or the positions of devices are changing, a period for the common beam training needs to be reduced, resulting in a substantial increase in the training overhead. 
In order to deal with this scenario, we use the \textit{dedicated beam} targeted for the users under high mobility on top of the existing common beam. Using the information on the target user's channel, we can improve the beam training efficiency by employing only a small number of beams (i.e., a few channel uses) optimized for the specific user. 
In the next subsections, we present the overall beam training protocol, dynamic channel characteristics, optimal beam design, and new channel estimation method.



\begin{figure} [t]
 \centering
  \includegraphics[width=4in]{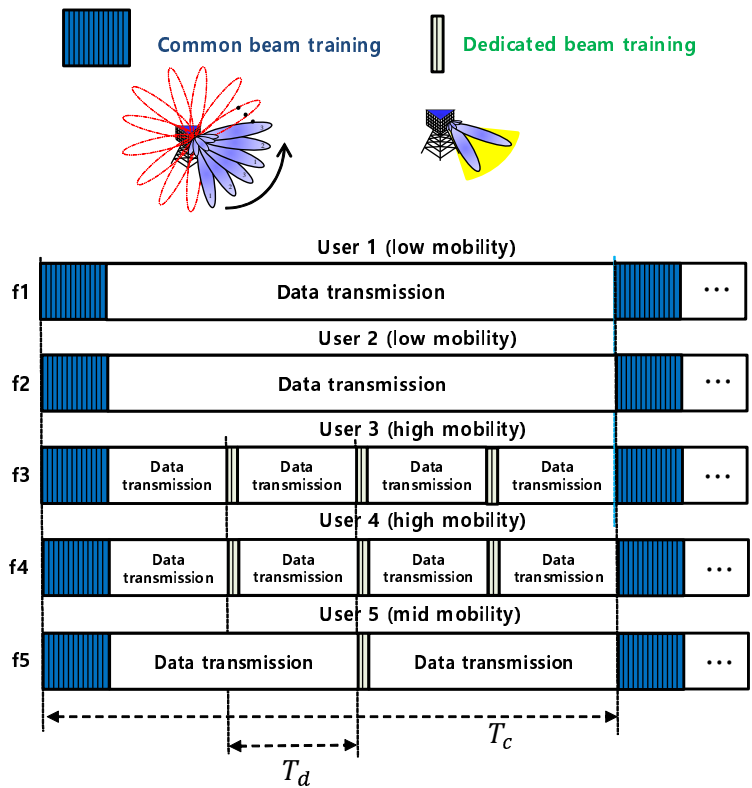}
 \\
 \caption{Frame structure of the proposed dedicated beam training. (for example of $N_d=2$ and $N_c=12$)} \label{fig:pro_scheme}
\end{figure}

\subsection{Description of Overall Beam Training Protocol} \label{subsec:overall}

Fig. \ref{fig:pro_scheme} depicts the proposed beam training protocol, where the five users are communicating with the base-station using different frequency resources.  For the users under mobility (e.g., user 3, 4, and 5 in Fig. \ref{fig:pro_scheme}), the base-station transmits the dedicated beams as well as the common beams.
As shown in Fig.~\ref{fig:pro_scheme}, the base-station sends $N_d$ training beams for the dedicated beam training where the number of dedicated beams $N_d$ is much smaller than the number of common beams $N_c$. In order to serve high mobility users, the dedicated beams are transmitted more frequently than the common beams, i.e., $T_d < T_c$. As shown in Fig.~\ref{fig:pro_scheme}, the periodicity of the dedicated beam can be adjusted depending on the extent of mobility. The proposed dedicated beam training involves the interaction between the base-station and the target user. Based on the received beams, the target user estimates the CSI and picks the optimal beams to be used by the base-station in the next beam training period.  The selected beam indices are fed back to the base-station. The proposed beam training procedure is repeated until the next common beam training period begins. 
Since the beamforming vectors are chosen based on the CSI of the user, the quality of channel estimation and beamforming can be maintained even with a few training beam transmissions. Note also that as illustrated in Fig. \ref{fig:pro_scheme}, the dedicated beam training utilizes only two channel uses for the beam transmission and employs the frequency resource allocated only to the target users.

\subsection{Dynamic mmWave Channel Characteristics} \label{subsec:statistical_model}

In this subsection, we describe the statistical characteristics of mmWave channel under the mobility scenario.  
Due to the physical constraint of the user's motion,  the current values of AoD and AoA do not deviate much from the previous ones.
Under the assumption that AoD and AoA are varying slowly and represented in the discrete space (see (\ref{eq:angular_model})), we model the AoD and AoA as the discrete state Markov process \cite{heathtr,he,sounding,mmwave_tracking1,mmwave_tracking2}.
The discrete state Markov process is described by the transition probability given by
\begin{align}
T^{(b)}(m|n) &= P\Big(\theta_{p,t,l}^{(b)} = -1+2\frac{m-1}{M_b} \bigg|\theta_{p,t-1,l}^{(b)}=-1+2\frac{n-1}{M_b}\Big) \nonumber \\
T^{(m)}(m|n) &= P\Big(\theta_{p,t,l}^{(m)} = -1+2\frac{m-1}{M_m} \bigg|\theta_{p,t-1,l}^{(m)}=-1+2\frac{n-1}{M_m}\Big). \label{eq:markov}
\end{align}
One simple example of the transition probability is
\begin{align} \label{eq:tmn}
T^{(b)}(m|n) = \frac{1}{C}\exp(-|m-n|^2/\sigma_{l}^2),
\end{align}
where $C$ is the normalization constant and $\sigma_l$ indicates the extent of mobility for the user. This transition probability decays exponentially with the difference between the present and past AoD.
Note that larger value of $\sigma_{p}^2$ implies higher mobility, and thus one can estimate the parameter $\sigma_{l}$ from the user's moving trajectory based on the maximum likelihood criteria.


\subsection{Optimal Beam Design}
During the dedicated beam training period, the base-station needs to adapt the beamforming vectors based on user's CSI. 
In our work, since the dedicated beam training is performed in the angular space, we use the channel information represented in terms of the  AoD and  AoA.
First, the proposed beam design selects $N_d$ beamforming vectors ($\mathbf{f}_{t,1},...,  \mathbf{f}_{t,N_d}$) {minimizing the channel estimation error of the target user} from the beam codebook $\mathcal{D}$.
We assume that the beam codebook  $\mathcal{D} = \left\{\mathbf{d}_1, ..., \mathbf{d}_R \right\}$ contains the $R$ pre-calculated beamforming vectors that correspond to the steering vectors for different directions. In our work, we only consider the analog beamforming vectors for simplicity but we can also consider additional beamforming vectors assuming more complex hardware such as hybrid beamforming architecture.
Then, the optimal beam selection problem can be formulated as
\begin{align} \label{eq:beamsel}
\left( \mathbf{f}_{t,1}^{*},...,  \mathbf{f}_{t,N_d}^{*} \right)  = \underset{\mathbf{f}_{t,1},...,  \mathbf{f}_{t,N_d} \in \mathcal{D}}{\arg\min}
E\left[\| \mathbf{h}_{p,t} - \hat{\mathbf{h}}_{p,t}\|_2^2 \right],
\end{align}
where $\hat{\mathbf{h}}_{p,t}$ is the estimate of $\mathbf{h}_{p,t}$. 
Since  both the combining vector $\mathbf{w}_{p,t}$  and the beamforming vector $\mathbf{f}_{t}$ affect the channel estimation performance, they should be optimized jointly. 
In order to simplify the optimization process, we decouple joint optimization step by using the ideal combining matrix $\mathbf{W}_{p,t} = \mathbf{A}_{p,t}^{(m)}$ in the system model and optimizing the cost function only with respect to the beamforming vectors $\mathbf{F}_t = [\mathbf{f}_{t,1},...,  \mathbf{f}_{t,N_d}]$. 
After the support $\mathcal{S}_h$ in $\mathbf{h}_{p,t}$,
(corresponding to the active AoD bins) is estimated, the channel gain can be estimated by the linear projection of $\mathbf{y}_{p,t}$ onto the subspace spanned by $\mathcal{S}_h$.
Since the channel gain is represented as a function of AoD, the channel estimation performance is determined by the AoD estimation accuracy. Thus, the beam selection problem can be reformulated as
\begin{align}  \label{eq:mse}
\mathbf{f}_{t,1}^{*},...,  \mathbf{f}_{t,N_d}^{*}
 = \underset{\mathbf{f}_{t,1},...,  \mathbf{f}_{t,N_d} \in \mathcal{D}}{\arg\min}
E\left[\sum_{l=1}^{L} \left| \theta^{(b)}_{p,t,l} - \hat{\theta}^{(b)}_{p,t,l} \right|^2  \right],
\end{align}
where $\hat{\theta}^{(b)}_{p,t,l}$ is the estimate of $\theta^{(b)}_{p,t,l}$.
Though the cost function in (\ref{eq:mse}) is a reasonable choice, the actual evaluation of the cost function  depends the channel estimation algorithm being used. 
To avoid the dependency of the cost function in (\ref{eq:mse}) on the algorithm selection, we use the analytic bound of AoD variance as a performance metric.
Specifically, we obtain the lower bound of $E\left[\sum_{l=1}^{L} \left| \theta^{(b)}_{p,t,l} - \hat{\theta}^{(b)}_{p,t,l} \right|^2  \right]$ using information inequality  \cite{poor}.  
Let $\underline{\theta} =  \left[\theta^{(b)}_{p,t,1},...,\theta^{(b)}_{p,t,L}, \alpha_{p,t,1}, ...,\alpha_{p,t,L} \right]^{T}$, then the Cramer-Rao Lower bound (CRLB) for the parameter $\underline{\theta}$ is given by
\begin{align}
&{\rm Cov}\left(\underline{\theta}\right) \succeq \mathbf{V}^{-1} \\
&(\mathbf{V})_{i,j} = E\left[\frac{\partial \log p(\mathbf{Y}_{p,t}|\underline{\theta})}{\partial (\underline{\theta})_i}\frac{\partial \log p(\mathbf{Y}_{p,t}|\underline{\theta})}{\partial (\underline{\theta})_j}\right],
\end{align}
where $\mathbf{V}$ is the Fisher's information matrix and  $ (\mathbf{V}^{-1})_{l,l}$ is the CRLB of  $\theta^{(b)}_{p,t,l}$ for $1 \leq l \leq L$.
In our setup, the CRLB is a function of the deterministic parameters $\theta^{(b)}_{p,t,1},...,\theta^{(b)}_{p,t,L}$ so that we cannot determine the CRLB without the knowledge of these parameters. 
As heuristic surrogate, we use the CRLB weighted with respect to the probability distribution $Pr(\theta^{(b)}_{p,t,1},...,\theta^{(b)}_{p,t,L})$ as a cost function, i.e.,
\begin{align}
C_{avg}(\mathbf{f}_{t,1},...,  \mathbf{f}_{t,N_d}) =& E_{\theta^{(b)}_{p,t,1},...,\theta^{(b)}_{p,t,L}} \left[ \sum_{l=1}^{L} [{\rm Cov}\left(\underline{\theta}\right)]_{l,l} \right]  \nonumber \\
 \geq&  E_{\theta^{(b)}_{p,t,1},...,\theta^{(b)}_{p,t,L}} \left[ \sum_{l=1}^{L} (\mathbf{V}^{-1})_{l,l} \right]  \nonumber \\
=&\sum_{\theta^{(b)}_{p,t,1},...,\theta^{(b)}_{p,t,L}} \bigg( \sum_{l=1}^{L} (\mathbf{V}^{-1})_{l,l} \bigg) Pr(\theta^{(b)}_{p,t,1},...,\theta^{(b)}_{p,t,L}) \nonumber \\
=&\sum_{m_1,...,m_L \in [1,M_b]} \bigg( \sum_{l=1}^{L} (\mathbf{V}^{-1})_{l,l}\bigg)  \prod_{i=1}^{L} Pr\Big(\theta^{(b)}_{p,t,i}= -1+2\frac{m_i-1}{M_b} \Big). \label{eq:weight}
\end{align}
where the AoDs  for different paths are assumed to be statistically independent in (\ref{eq:weight}). The distribution of the AoD is obtained by the proposed channel estimator, which will be described later. Specifically, based on the training beams received during the previous beam training periods, the proposed channel estimator predicts the distribution of the AoD for the next beam training period. By doing so, we can account for the uncertainty on the AoD estimate in our beam selection strategy. 
Since the cost function is expressed as a function of $N_d$ beamforming vectors $\mathbf{f}_{t,1},...,  \mathbf{f}_{t,N_d}$, the beam selection rule can be readily expressed as 
\begin{align}
\left(\mathbf{f}_{t,1}^{*},...,  \mathbf{f}_{t,N_d}^{*}\right) = \underset{\mathbf{f}_{t,1},...,  \mathbf{f}_{t,N_d} \in \mathcal{D}}{\arg\min}
C_{avg}(\mathbf{f}_{t,1},...,  \mathbf{f}_{t,N_d}).
 \label{eq:bsopt}
\end{align}

\subsection{Beam Selection with Single Path Scenario \label{subsec:selection}}

\begin{figure} [t]
 \centering
 \subfigure[]{
 \includegraphics[width=3in]{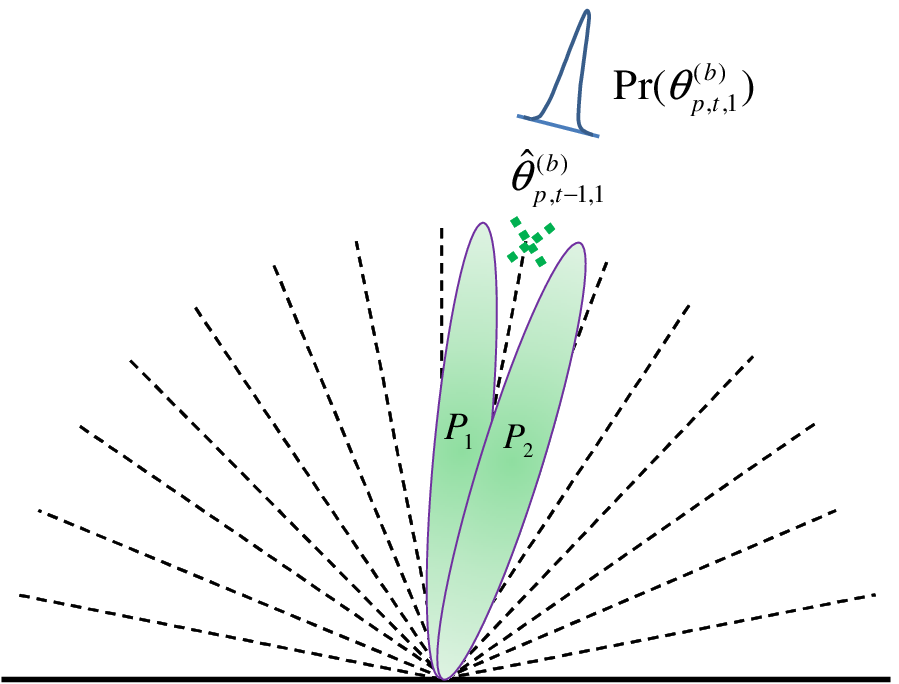}
 }
 \subfigure[]{
 \includegraphics[width=3in]{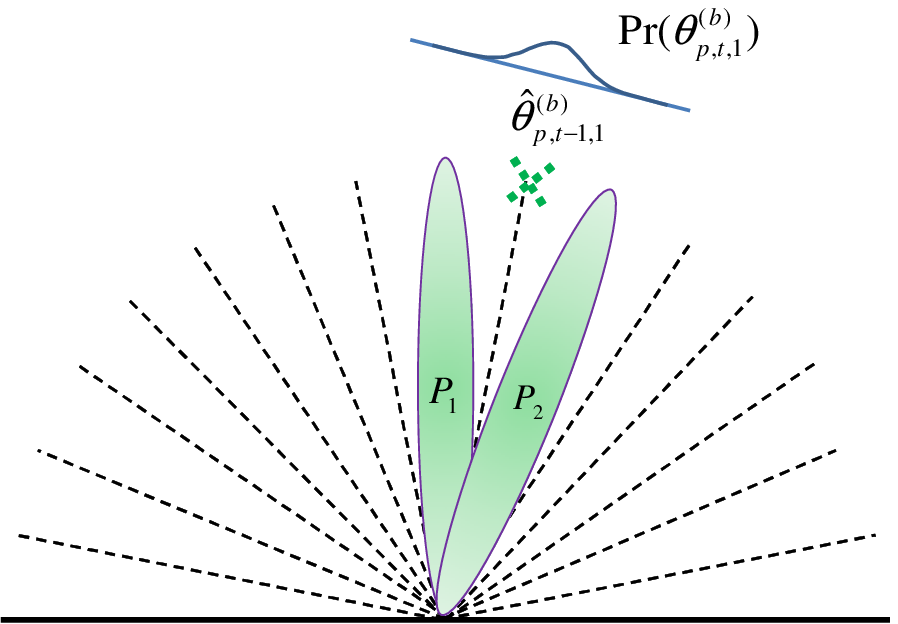}
 }
 \\
 \caption{Example of the proposed beam selection for different AoD distributions.} \label{fig:search_range}
\end{figure}

First, we consider the beam selection rule for the single path scenario (i.e., $L=1$) where the energy of single path is predominant.
In this scenario, the beam selection rule in (\ref{eq:mse})  can be simplified to
\begin{align}
\left(\mathbf{f}_{t,1}^{*},...,  \mathbf{f}_{t,N_d}^{*}\right) = \underset{\mathbf{f}_{t,1},...,  \mathbf{f}_{t,N_d} \in \mathcal{D} }{\arg\min}
 E\left| \theta^{(b)}_{p,t,1} - \hat{\theta}^{(b)}_{p,t,1} \right|^2.
\end{align}
Also, since $L=1$, the received signal vector $\mathbf{Y}_{p,t}$ in (\ref{eq:an}) is simplified to
\begin{align} \label{eq:alpha}
\mathbf{Y}_{p,t} = \alpha_{p,t,1} \mathbf{W}_t^{H} \mathbf{a}^{(m)}(\theta^{(m)}_{p,t,1})\big(\mathbf{a}^{(b)}(\theta^{(b)}_{p,t,1})\big)^{H} \mathbf{F}_t + \mathbf{N}'_t.
\end{align}
Using the ideal combining matrix  $\mathbf{W}_t = \mathbf{a}^{(m)}(\theta^{(m)}_{p,1,t})$, we have
\begin{align} \label{eq:ypt}
\mathbf{Y}_{p,t} = \alpha_{p,t,1} \left(\mathbf{a}^{(b)}(\theta^{(b)}_{p,t,1})\right)^{H} \mathbf{F}_t + \mathbf{N}'_t.
\end{align}
Since  $\mathbf{Y}_{p,t}$ is a row vector, we can rewrite (\ref{eq:ypt}) as
\begin{align}
\mathbf{y}_{p,t} = \mathbf{Y}_{p,t}^{T} =  \mathbf{F}_t^{T} \mathbf{a}^{(b)}(\theta^{(b)}_{p,t,1})^*\alpha_{p,t,1}  + \mathbf{n}_{p,t}, \label{eq:l1}
\end{align}
and one can show that the CRLB of $\theta^{(b)}_{p,t,1}$ is given by (see the detailed derivation in Appendix A)
\begin{align}
  Var(\theta^{(b)}_{p,t,1})  &\geq   ( \mathbf{V}^{-1})_{3,3} \nonumber \\
  &= C(\mathbf{f}_{t,1},...,  \mathbf{f}_{t,N_d}) \nonumber \\
  & = \frac{\sigma^2}{\alpha_{p,t,1}^2} \left[
  \left\|\mathbf{F}_{t}^{T} \frac{ \partial \mathbf{a}^{(b)}(\theta_{p,t,1}^{(b)})^{*}}{\partial \theta_{p,t,1}^{(b)}}\right\|^2
  - \frac{\left\| \frac{\partial \mathbf{a}^{(b)}(\theta_{p,t,1}^{(b)})^{T}}{\partial \theta_{p,t,1}^{(b)}} \mathbf{F}_{t}^{*}\mathbf{F}_{t}^{T} \mathbf{a}^{(b)}(\theta_{p,t,1}^{(b)})^{*} \right\|^2
  }{\left\|\mathbf{F}_{t}^{T} \mathbf{a}^{(b)}(\theta_{p,t,1}^{(b)})^{*}\right\|^2}
   \right]^{-1} .
\label{eq:CRLB}
\end{align}
Note from (\ref{eq:CRLB}) that $\alpha_{p,t,1}$ does not affect the beamforming matrix $\mathbf{F}_{t}$ minimizing the lower bound.
By taking similar step to (\ref{eq:weight}), the average cost function can be obtained by weighting $C(\mathbf{f}_{t,1},...,  \mathbf{f}_{t,N_d})$ with the distribution of $\theta^{(b)}_{p,t,1}$, i.e., 
\begin{align}
 C_{avg}(\mathbf{f}_{t,1},...,  \mathbf{f}_{t,N_d}) 
& = \sum_{m=1}^{M_b} C(\mathbf{f}_{t,1},...,  \mathbf{f}_{t,N_d}) Pr\left(\theta^{(b)}_{p,t,1}= -1+2\frac{m-1}{M_b} \right).
\end{align}
Then, we search for $N_d$ beam indices from the beam codebook $\mathcal{D}$ that minimize the cost function  $C_{avg}(\mathbf{f}_{t,1},...,  \mathbf{f}_{t,N_d})$.
Although this process is combinatoric in nature and thus computationally burdensome,  the computational complexity can be reduced significantly by considering only a small number of beams.
In our work, we consider dual beam transmission ($N_d=2$), which offers small search complexity as well as low training overhead. In fact, we will demonstrate from simulations that the performance of the proposed method using two training beams is  comparable to the common beam training using $N_c=32$ beams. In addition, we will also show that using a single beam does not provide satisfactory performance.
When $N_d = 2$, we search for two  beam indices $(i,j)$ from $\mathcal{D}$ as
\begin{align}
(i^*, j^*) = \underset{(i,j) \in \mathcal{D}}{\arg\min} \  C_{avg}(\mathbf{f}_{t,1}=\mathbf{d}_i,\mathbf{f}_{t,2}=\mathbf{d}_j).
\end{align}
To find out $\left(i^*,j^*\right)$, we need to evaluate $\binom{M_b}{2}$ combinations of all beam indices. 
When the AoD distribution is symmetric, we observe that the directions of the optimal beam pair are also symmetric with respect to the center of the distribution. (see the illustration in Fig.~\ref{fig:search_range}.)
Exploiting this symmetry, we only have to determine the angle formed by two beamforming vectors. Assuming that the codebook contains the beamforming vectors with equally spaced directions and the directions of the beamforming vectors $\mathbf{d}_{i_0+\delta}$ and $\mathbf{d}_{i_0-\delta}$ are symmetric with respect to that of $\mathbf{d}_{i_0}$, two dimensional grid search can be reduced to the one dimensional search as
\begin{align}
(i_0+\delta^*,i_0-\delta^*) &= \underset{    i_0+\delta, i_0-\delta \in \mathcal{D}}{\arg\min} C_{avg}(\mathbf{f}_{t,1}=\mathbf{d}_{i_0+\delta},\mathbf{f}_{t,2}=\mathbf{d}_{i_0-\delta}),
\end{align}
where $i_0$ is the beam index corresponding to the center of the AoD distribution.
Note that the beam index parameter $\delta^*$ corresponds to the optimal angle between two selected training beams.  (see Fig.~\ref{fig:search_range}.)

\subsection{Beam Selection With Multi-path Scenario}

When there are multiple paths, we need to select beams based on the cost metric defined over the set of parameters $\underline{\theta} = [\alpha_{p,t,1},\alpha_{p,t,1}^*,...,\alpha_{p,t,L},\alpha_{p,t,L}^*, \theta^{(b)}_{p,t,1},..,\theta^{(b)}_{p,t,L} ]$.
As the number of parameters to be considered increases, the derivation of the CRLB would be cumbersome and also the optimization process requires huge computational complexity. To avoid this hassle, we use a simple beam selection method, which uses the dual beams optimized for each path. That is, we separately optimize the dual beams for each path under the assumption of $L=1$ and use the optimized beams for all $L$ paths.
Thus, the base-station sends $2L$ beams in total. Since the number of paths $L$ is small in mmWave environments, the training resources for the dedicated beam training is much smaller than the common beam training (i.e., $2L \ll N_c$). Since our channel estimator is designed to provide the
distribution of AoD for each path, (i.e., $Pr( \theta^{(b)}_{p,t,1}),...,  Pr(\theta^{(b)}_{p,t,L})$), the optimal beam spacing $\delta^{*}$  is obtained by running the one dimensional search algorithm for each path.
Though such proposed per-path beam selection seems to be a bit heuristic, our numerical evaluation shows that it offers good system performance with reasonably small training overhead.

\subsection{Sequential Greedy Channel Estimation for Dedicated Beam Training} \label{subsec:signal_model}

In order to estimate the channel $\mathbf{H}_{p,t}$ in mmWave communication systems, we first need to estimate the channel parameters $\alpha_{p,t,1},..., \alpha_{p,t,L}$, $\theta^{(b)}_{p,t,1},..., \theta^{(b)}_{p,t,L}$, and $\theta^{(m)}_{p,t,1},..., \theta^{(m)}_{p,t,L}$
from the received vector $ \mathbf{y}_{p,t}$.
Under the quantized channel model, ($\mathbf{y}_{p,t}= \mathbf{\Phi}_{p,t} \mathbf{h}_{p,t} + \mathbf{n}_{p,t}$), the AoD and AoA parameters correspond to the support of the channel vector $\mathbf{h}_{p,t}$ (i.e., the set of indices of nonzero elements in $\mathbf{h}_{p,t}$). 
In this subsection, we propose a greedy sparse channel estimation algorithm that jointly estimates the support and the amplitude of $\mathbf{h}_{p,t}$ from $ \mathbf{y}_{p,t}$.
Note that the proposed channel estimator can sequentially generate the distribution of  the current AoD parameter $Pr(\theta^{(b)}_{p,t,l})$  using the measurement vectors $\mathbf{y}_{p,1},...,\mathbf{y}_{p,t-1}$ acquired until the $(t-1)$th beam training period.

Let $\mathcal{S}_{p,t}=\{s_{p,t,1},...,s_{p,t,L} \}$ be the support of the channel vector  $\mathbf{h}_{p,t}$, where the support index $s_{p,t,l}$ corresponds to the pair of the AoD $\theta^{(b)}_{p,t,l}$ and the AoA $\theta^{(m)}_{p,t,l}$.  Then, the received vector $\mathbf{y}_{p,t}$ can be expressed as
\begin{align}
\mathbf{y}_{p,t}=(\mathbf{\Phi}_{p,t})_{\mathcal{S}_{p,t}} \mathbf{g}_{p,t} + \mathbf{n}_{p,t}.
\end{align}
where  $\mathbf{g}_{p,t}$ is the $L \times 1$  vector containing nonzero gains in $\mathbf{h}_{p,t}$.
Recall that $(\mathbf{X})_{\Omega}$ is the submatrix of $\mathbf{X}$ that contains columns indexed by $\Omega$
\footnote{For example, 
$\left(\begin{bmatrix}1&2&3&4\\ 4&3&2&1 \end{bmatrix} \right)_{\{2,4\}} = \begin{bmatrix}2&4\\ 3&1 \end{bmatrix}$.
}.
We assume that $\mathbf{g}_{p,t}$ is the Gaussian random vector $CN(0,\lambda\mathbf{I})$. Since the channel vector $\mathbf{h}_{p,t}$ is completely determined by two parameters, viz., the support $\mathcal{S}_{p,t}$ and the magnitude of the gain $\mathbf{g}_{p,t}$, the channel estimate can be obtained by finding the joint estimate of  $\mathcal{S}_{p,t}$ and the amplitude $\mathbf{g}_{p,t}$. The joint maximum a posteriori (MAP) estimate of $\mathcal{S}_{p,t}$ and  $\mathbf{g}_{p,t}$ is given by
\begin{align}
\left(\hat{\mathcal{S}}_{p,t}, \hat{\mathbf{g}}_{p,t} \right) = \underset{\mathcal{S}_{p,t}, \mathbf{g}_{p,t}}{\arg \max} \ln Pr\left(\mathcal{S}_{p,t}, \mathbf{g}_{p,t}| \mathbf{y}_{p,1},..., \mathbf{y}_{p,t}\right).
\end{align}
Using $Pr(\mathcal{S}_{p,t}, \mathbf{g}_{p,t}|\mathcal{Y}_{p,t}) = Pr(\mathcal{S}_{p,t}|\mathcal{Y}_{p,t}) Pr(\mathbf{g}_{p,t}| \mathcal{S}_{p,t},\mathcal{Y}_{p,t})$ and denoting $\mathcal{Y}_{p,t} = \{\mathbf{y}_{p,1},..., \mathbf{y}_{p,t} \}$, the MAP estimate $\hat{\mathcal{S}}_{p,t}$ of the support can be obtained from
\begin{align}
\hat{\mathcal{S}}_{p,t}  = & \underset{\mathcal{S}_{p,t}}{\arg \max}\bigg[ \max_{\mathbf{g}_{p,t}} \bigg( \ln Pr(\mathcal{S}_{p,t}|\mathcal{Y}_{p,t}) + \ln Pr(\mathbf{g}_{p,t}| \mathcal{S}_{p,t},\mathcal{Y}_{p,t}) \bigg) \bigg] \\
 =& \underset{\mathcal{S}_{p,t}}{\arg \max} \bigg[ \ln Pr(\mathcal{S}_{p,t}|\mathcal{Y}_{p,t})  + \max_{\mathbf{g}_{p,t}}  \bigg( \ln Pr(\mathbf{g}_{p,t}| \mathcal{S}_{p,t},\mathcal{Y}_{p,t}) \bigg) \bigg].   \label{eq:spt}
\end{align}
Note that the channel amplitude vector $\mathbf{g}_{p,t}$ is Gaussian distributed when the support  $\mathcal{S}_{p,t}$ is given, i.e., $Pr(\mathbf{g}_{p,t}| \mathcal{S}_{p,t},\mathcal{Y}_{p,t}) \sim \mathcal{CN}(\bar{\mathbf{g}}_{p,t},P_{p,t})$ where
\begin{align*}
\bar{\mathbf{g}}_{p,t} &=\left((\mathbf{\Phi}_{p,t})_{\mathcal{S}_{p,t}}^{H} (\mathbf{\Phi}_{p,t})_{\mathcal{S}_{p,t}} + \frac{\sigma_n^2}{\lambda} I\right)^{-1} (\mathbf{\Phi}_{p,t})_{\mathcal{S}_{p,t}}^{H}  \mathbf{y}_{p,t}\\
P_{p,t} &=\sigma_n^2 \left((\mathbf{\Phi}_{p,t})_{\mathcal{S}_{p,t}}^{H} (\mathbf{\Phi}_{p,t})_{\mathcal{S}_{p,t}} + \frac{\sigma_n^2}{\lambda} I\right)^{-1}.
\end{align*}
Since $\bar{\mathbf{g}}_{p,t}$ is the maximizer of $Pr(\mathbf{g}_{p,t}| \mathcal{S}_{p,t},\mathcal{Y}_{p,t})$ in (\ref{eq:spt}), we have
\begin{align}
\hat{\mathcal{S}}_{p,t} =& \underset{\mathcal{S}_{p,t}}{\arg \max} \left[ \ln Pr(\mathcal{S}_{p,t}|\mathcal{Y}_{p,t}) - \ln \left( \left|\pi P_{p,t}\right| \right) \right].  \label{eq:spt2}
\end{align}
One can also show that the term $\ln Pr(\mathcal{S}_{p,t}|\mathcal{Y}_{p,t})$ in (\ref{eq:spt2}) can be expressed with respect to $Pr(\mathcal{S}_{p,t-1}|\mathcal{Y}_{p,t-1})$ in a recursive form as
\begin{align}
\ln  Pr(\mathcal{S}_{p,t}|\mathcal{Y}_{p,t}) =& \ln Pr(\mathcal{S}_{p,t}|\mathcal{Y}_{p,t-1},\mathbf{y}_{p,t}) \nonumber \\
= & \ln Pr(\mathbf{y}_{p,t}\left|\mathcal{S}_{p,t}, \mathcal{Y}_{p,t-1}) + \ln Pr(\mathcal{S}_{p,t}\right|\mathcal{Y}_{p,t-1}) +C \nonumber \\
= &- \ln \left|\pi \left( \sigma_n^2 I+\lambda(\mathbf{\Phi}_{p,t})_{\mathcal{S}_{p,t}}^{H} (\mathbf{\Phi}_{p,t})_{\mathcal{S}_{p,t}}\right)\right|+ \mathbf{y}_{p,t}^{H} (\mathbf{\Phi}_{p,t})_{\mathcal{S}_{p,t}} 
P_{p,t}
(\mathbf{\Phi}_{p,t})_{\mathcal{S}_{p,t}}^{H}  \mathbf{y}_{p,t}  \nonumber \\
 &  + \ln \sum_{\mathcal{S}_{p,t-1}} Pr(\mathcal{S}_{p,t}|\mathcal{S}_{p,t-1}) Pr(\mathcal{S}_{p,t-1}|\mathcal{Y}_{p,t-1})  +C',   \label{eq:recur}
\end{align}
where $C$ and $C'$ are the terms unrelated to $\mathcal{S}_{p,t}$.  Note that the conditional distribution $Pr(\mathcal{S}_{p,t}|\mathcal{S}_{p,t-1})$    is determined by the discrete-state Markov process described in (\ref{eq:markov}).
Plugging (\ref{eq:recur}) into (\ref{eq:spt2}), we obtain the expression for the objective function as
\begin{align}
\psi(\mathcal{S}_{p,t}) =& - \ln \left|\pi \left( \sigma_n^2 I+\lambda(\mathbf{\Phi}_{p,t})_{\mathcal{S}_{p,t}}^{H} (\mathbf{\Phi}_{p,t})_{\mathcal{S}_{p,t}}\right)\right| 
+ \mathbf{y}_{p,t}^{H} (\mathbf{\Phi}_{p,t})_{\mathcal{S}_{p,t}} P_{p,t}  (\mathbf{\Phi}_{p,t})_{\mathcal{S}_{p,t}}^{H}  \mathbf{y}_{p,t}  \nonumber \\
 &  + \ln \sum_{\mathcal{S}_{p,t-1}} Pr(\mathcal{S}_{p,t}|\mathcal{S}_{p,t-1}) Pr(\mathcal{S}_{p,t-1}|\mathcal{Y}_{p,t-1})   -  \ln \left( \left|\pi P_{p,t}\right| \right). \label{eq:objs} 
\end{align}
To find out the MAP estimate of the support, we need to evaluate (\ref{eq:objs}) for all possible combinations of $\mathcal{S}_{p,t}$.
Since the  search complexity growth exponentially with $L$, this option is infeasible in practice.
\begin{algorithm} [t] \label{alg:proposed}
	\caption{Proposed greedy recovery algorithm}
	\textbf{Initialize : }$\mathbf{r}^{(0)} = \mathbf{y}_{p,t},  \Omega_0 = \{\}$ \\
	\textbf{Output : } support set $\Omega$, signal amplitude $\hat{x}_{\Omega}$, AoD distribution, $Pr(\mathcal{S}_{p,t+1}= \{\omega\}|\mathcal{Y}_{p,t})$
	\begin{algorithmic}[1]
		\For {$l=1$ to $L$$...$}
		\State Select the index $\omega$ that leads to the largest objective function:
\begin{align*}
 \omega^{(i)}= \arg\underset{\omega}{\max} ~\psi(\omega)
 \end{align*}
%
		\State Update the support set $\Omega_i$
\begin{align*}
\Omega_i = \Omega_{i-1}\cup   \omega^{(i)}
\end{align*}

\State Update the signal amplitude set $\mathbf{x}_{\Omega_i}$ and the residual signal $\mathbf{r}_t^{(l)}$
		\begin{align*}
		 &\mathbf{x}_{\Omega_i}  = \left((\mathbf{\Phi}_{p,t})_{\Omega_{i}}^{H} (\mathbf{\Phi}_{p,t})_{\Omega_{i}} + \frac{\sigma_n^2}{\lambda} I\right)^{-1} (\mathbf{\Phi}_{p,t})_{\Omega_{i}}^{H}  \mathbf{y}_{p,t}
		\\ &\mathbf{r}_t^{(l)} =  \mathbf{y}_{p,t} - (\mathbf{\Phi}_{p,t})_{\Omega_{i}} \mathbf{x}_{\Omega_i}
		\end{align*}
		\State Update the \emph{a posteriori} distribution using (\ref{eq:fome}).

			\EndFor
		\State Calculate the AoD distribution for the next beam training period using (\ref{eq:om})
	\end{algorithmic}
\end{algorithm}

In this work, we propose a computationally efficient greedy algorithm that finds the candidate element of $\mathcal{S}_{p,t}$ in an iterative fashion. 
Adopting the greedy strategy, we propose a cost function to find a predominant support index. Let $\Omega_i$ be the set of the support indices found by the $i$th iteration, then we compute the residual signal $\mathbf{r}^{(i)}$ for the $i$th iteration by subtracting the contribution of the already found support elements in $\Omega_{i}$ from the observation $\mathbf{r}^{(i)}$. That is,
\begin{align}
\mathbf{r}^{(i)} & = \mathbf{y}_{p,t} - (\mathbf{\Phi}_{p,t})_{\Omega_{i}}\mathbf{x}_{\Omega_i}
\end{align}
where
\begin{align*}
\mathbf{x}_{\Omega_i}  = \left((\mathbf{\Phi}_{p,t})_{\Omega_{i}}^{H} (\mathbf{\Phi}_{p,t})_{\Omega_{i}} + \frac{\sigma_n^2}{\lambda} I\right)^{-1} (\mathbf{\Phi}_{p,t})_{\Omega_{i}}^{H}  \mathbf{y}_{p,t}.
\end{align*}
Note that we assume that the effect of the support indices in $\Omega_{i}$ is perfectly subtracted from the residual signal $\mathbf{r}^{(i)}$.
Since there exist a dominant support element  in the residual $\mathbf{r}^{(i)}$, we aim to find it from the residual $\mathbf{r}^{(i)}$.
Assuming  the single support candidate $\omega \notin  \Omega_{i-1}$,  the objective function in (\ref{eq:objs}) becomes
\begin{align} 
 \psi(\omega) =& - \ln \left|\pi \left( \sigma_n^2 I+\lambda(\mathbf{\Phi}_{p,t})_{\omega}^{H}(\mathbf{\Phi}_{p,t})_{\omega}\right)\right|  + \frac{1}{\sigma_n^2} \frac{| (\mathbf{\Phi}_{p,t})_{\omega}^{H} \mathbf{r}^{(i-1)}|^2}{\|(\mathbf{\Phi}_{p,t})_{\omega}\|^2 + \frac{\sigma_n^2}{\lambda}} \nonumber \\
&+ \ln \sum_{v} Pr\left(s_{p,t}^{(i)}=\omega|s_{p,t-1}^{(i)}=v\right) Pr\left(s_{p,t-1}^{(i)}=v|\mathcal{Y}_{p,t-1}\right) - \pi \frac{\sigma_n^2}{\|(\mathbf{\Phi}_{p,t})_{\omega}\|^2 + \frac{\sigma_n^2}{\lambda}I}, \label{eq:psi}
\end{align}
where $s_{p,t}^{(i)}$ is the index of the support not in $\Omega_{i-1}$. 
Note that while the first two terms of (\ref{eq:psi}) are  also used in the cost function of the original OMP algorithm, 
the last two terms are produced to account for temporal correlations of the support elements.
Note that $Pr\left(s_{p,t}^{(i)}=\omega|s_{p,t-1}^{(i)}=v\right)$ can be given by the transition probability in (\ref{eq:tmn}), i.e., 
$\frac{1}{C}\exp(-|v-\omega|^2/\sigma_l^2)$.
In a similar way as in (\ref{eq:recur}), the term $Pr\left(s_{p,t-1}^{(i)}=v|\mathcal{Y}_{p,t-1}\right)$ in (\ref{eq:psi}) can be recursively calculated as
\begin{align}
\ln Pr(s_{p,t}^{(i)}=\omega|\mathcal{Y}_{p,t}) 
=& \frac{1}{\sigma_n^2} \frac{| (\mathbf{\Phi}_{p,t})_{\omega}^{H} \mathbf{r}^{(i-1)}|^2}{\|(\mathbf{\Phi}_{p,t})_{\omega}\|^2 + \frac{\sigma_n^2}{\lambda}} - \ln \left|\pi \left( \sigma_n^2 I+\lambda(\mathbf{\Phi}_{p,t})_{\omega}^{H} (\mathbf{\Phi}_{p,t})_{\omega}\right)\right| \nonumber \\
&+ \ln \sum_{v} Pr\left(s_{p,t}^{(i)}=\omega|s_{p,t-1}^{(i)}=v\right)  Pr\left(s_{p,t-1}^{(i)}=v|\mathcal{Y}_{p,t-1}\right). \label{eq:fome}
\end{align}
Further, using the result in (\ref{eq:fome}), the distribution of the AoD for the next beam training period can be obtained as
\begin{align} \label{eq:om}
Pr(s_{p,t+1}^{(i)}=\omega|\mathcal{Y}_{p,t}) =  \sum_{v} Pr(s_{p,t+1}^{(i)}=\omega|s_{p,t}^{(i)}=v) Pr(s_{p,t}^{(i)}=v|\mathcal{Y}_{p,t}).
\end{align}
This distribution is used to perform the beam selection for the next beam training period. 
We summarize the proposed greedy algorithm in Algorithm 2.

 \begin{figure} [!t]
	\centering
	\subfigure[]{
		\includegraphics[width=3.5in]{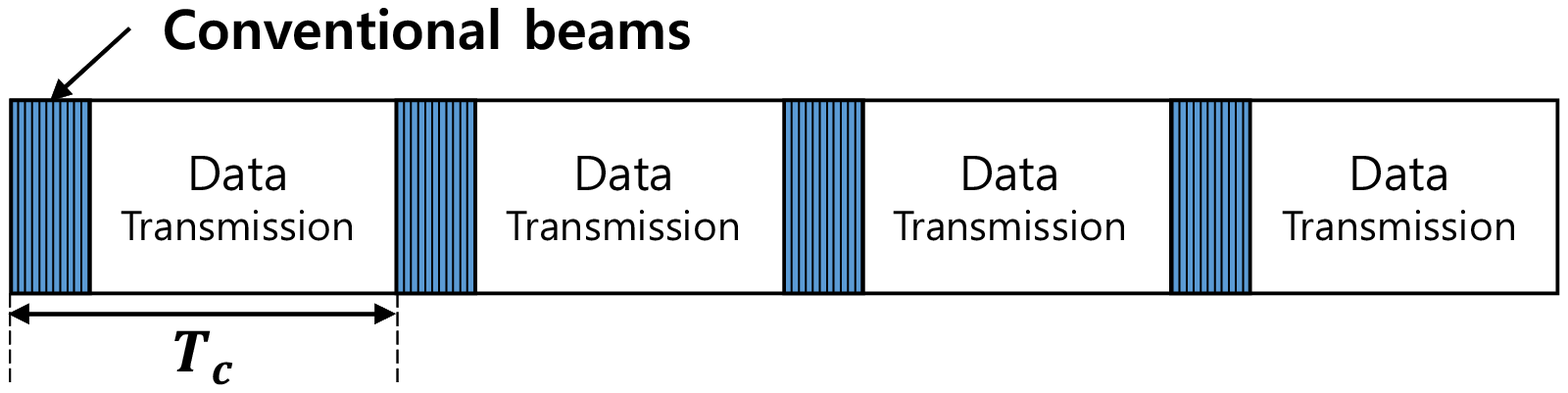}
	}
%
	\subfigure[]{
		\includegraphics[width=3.5in]{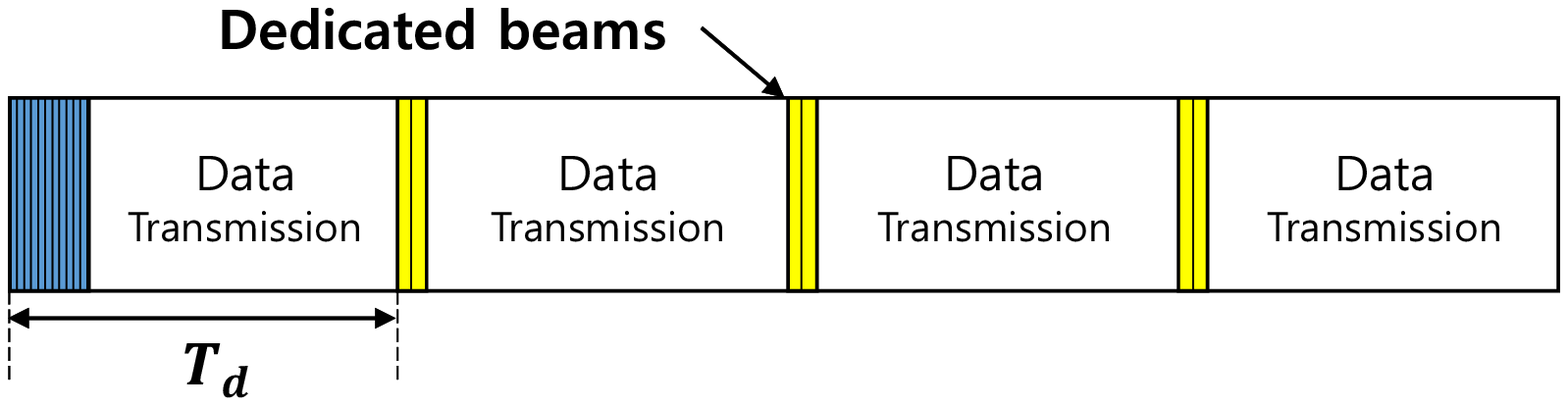}
	}
	\caption{Frame structure of (a) Conventional beam training and (b) dedicated beam training.}
	
	\label{fig:frame}
\end{figure}


\section{Simulation Results and Discussions\label{sec:simul} }

\subsection{Simulation Setup}
In this section, we evaluate the performance of the proposed dedicated beam training technique.
In our simulations, we consider 28GHz frequency band and uniform linear arrays (ULAs) antenna whose adjacent antenna elements are spaced by the half wavelength. 
The numbers of the antenna elements in the base-station and the user are set to $N_{b}=N_{m}=32$. Fig. \ref{fig:frame} (a) and (b) present the frame structures for the conventional common beam training and the proposed dedicated training used in our simulations. 
Following 5G NR standard \cite{5g_standard}, we set the symbol duration to 4.46$\mu$s.
For both common and dedicated beam training, the beam cycling ($N_c=32$ beam transmissions) is performed every 1000 symbols. In the conventional beam training, the beam cycling is performed every $T_c(<1000)$ symbols. In the proposed beam training protocol, on the other hand, the dedicated beam training is performed every $T_d$ symbols after the initial beam cycling.
 For the dedicated beam training, we use the dual beams (see Section \ref{sec:proposed_scheme}.D for details).
 The rest of the symbol slots  (i.e., not used for training beams) are allocated for data transmission.
 The data is modulated by the binary phase shift keying (BPSK) modulation. In order to focus on the performance of beam training, we use the optimal precoding scheme, which calculates the matrix $V$ from the singular value decomposition of the channel matrix $\hat{\mathbf{H}}_{p,t}= U \Sigma V^H$ and precode the data symbols by multiplying the matrix $V$ to data symbols. 
 In the simulations, the channel matrix is generated according to the model described in Section \ref{sec:sys_model}.B. 
We assume that the AoD and AoA change every symbol period according to the distribution $T^{(b)}(m|n) = \frac{1}{C}\exp(-|m-n|^2/\beta^2)$.  
 Note that $\sigma_l^2$ in (\ref{eq:tmn}) can be converted from $\beta^2$ as $\sigma_l ^2= {T_c}{\beta^2}$.
We also assume that the channel gain changes according to auto-regressive (AR) process given by  
\begin{align}
 \alpha_{n} =\rho \alpha_{n-1}+v_{n}\sqrt{1-\rho^2} 
 \end{align}
 where $v_{n} \sim N(0,1)$ and $\rho = 0.999$. Although the channel variation seems to be small, we see that $\rho = 0.999$ leads to significant change in channel gain across adjacent beam training periods. 
 The resolution of the quantized AoD and AoA representation is 
 $M_b=M_m=128$.    
The beam codebook contains 128 beamforming vectors whose angles are uniformly spaced between $[-\pi/2, \pi/2]$.
We generate a total of $10^6$ symbols to evaluate the bit error rate (BER). We define the training overhead as the ratio of the number of symbol slots used for training to the total number of symbols.

\begin{figure} [!t]
	\centering
	\includegraphics[width=3.8in,height=3.2in]{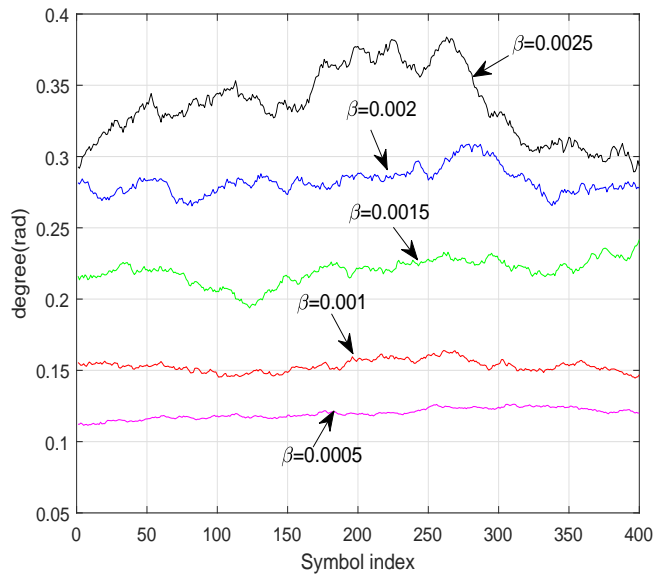}
	\\
	\caption{Temporal variation of AoD or AoA for several values of $\beta$.} \label{fig:angle_variation}
\end{figure}



\subsection{Simulation Results} \label{subsec:ber_perf}

\begin{figure} [t] 
	\centering
	\includegraphics[width=3.8in,height=3.2in]{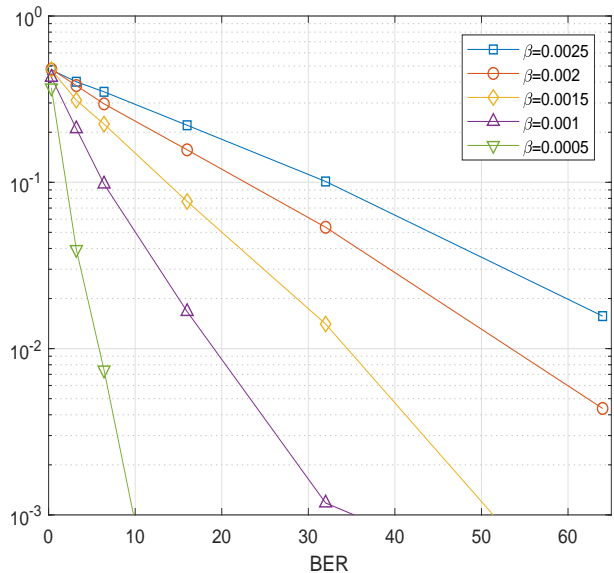}
	\caption{Plot of BER versus training overhead for the conventional beam cycling.} \label{fig:overhead_conv1}
\end{figure}




First, we investigate how fast the AoD (or AoA) changes in time  with respect to the parameter $\beta$. Fig. \ref{fig:angle_variation} shows the sample path of AoD (or AoA) in our channel model for different values of $\beta$. As $\beta$ increases, the extent of mobility of users increases and thus AoD (or AoA) exhibits higher variations. In order to find how parameter $\beta$ is related to the actual speed of a moving user, we generate the moving trajectory samples based on the second-order differential mobility model in \cite{mobility_model} for the given speed of the user. Then, we obtain the maximum likelihood (ML) estimate of $\beta$ from the trajectory samples. 
We see that $\beta = 0.5\times10^{-3}$ corresponds to the user moving at the speed of $2 m/s$ $(=7.2 km/h)$ at 100 meter away from the base-station.
Note that $\beta$ scales linearly with the speed of the user.

Fig. \ref{fig:overhead_conv1} shows performance of the conventional beam cycling under mobility scenarios. We plot BER performance as a function of the training overhead for different mobility levels. We consider the multi-path channel with  $L=3$ and the signal to noise ratio (SNR) is set to $15$ dB. 
When the training period $T_c$ is set to $1000$, 
the common beam training transmits $N_c=32$ training beams every 1000 symbols, resulting in the training overhead of $3.2$\%. As the training period $T_c$ is reduced to $500, 200$ and $100$, the training overhead increases up to $6.4$\%, $12.8$\%, and $32$\%, respectively. As expected, the BER performance degrades as the extent of mobility for the users increases. 
To achieve better performance, the common beam training should be performed frequently, resulting in the substantial increase in training overhead. For example, at the mobility of $\beta = 0.002$, the system should use more than 50\% training overhead to achieve $10^{-2}$ BER. This results clearly demonstrate that the conventional beam training is inefficient for the mobility scenarios.

\begin{figure} [!t]  
	\centering
	\subfigure[]{
		\includegraphics[width=3.8in,height=3.2in]{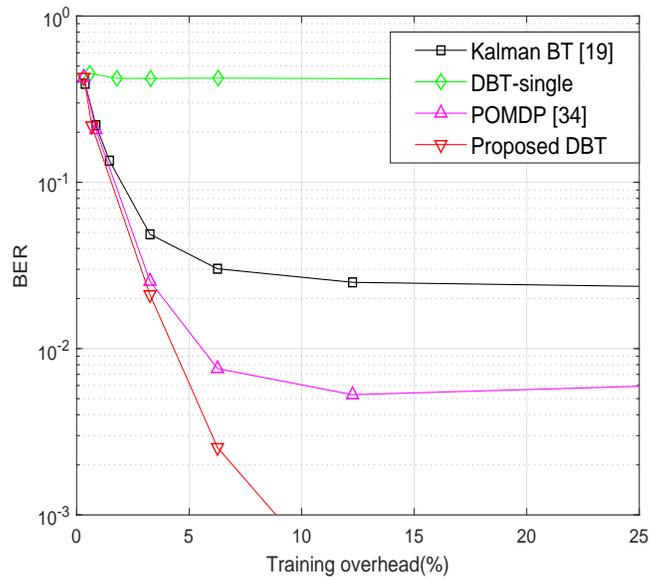}
	} 
	\subfigure[]{
		\includegraphics[width=3.8in,height=3.2in]{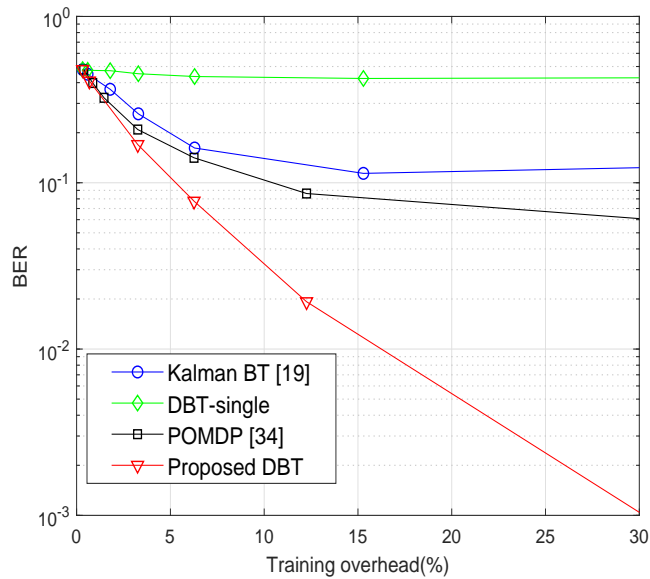}
	}
     \caption{Plot of BER versus training overhead for the dedicated beam training for (a) $\beta=0.001$ and (b) $\beta=0.002$. }
    \label{fig:overhead_new}
\end{figure}

We next evaluate the performance of several beam training strategies. 
The following beam training protocols are considered in our experiments
\begin{enumerate}
\item  Conventional Kalman-based beam tracking \cite{heathtr} (Kalman BT): After the initial beam cycling, the single beam is transmitted with period $T_d$. The AoD and AoA are estimated by the extended Kalman filter. 
\item POMDP-based beam sequence design \cite{sung}:  $N_d$ beams are designed under the POMDP framework. Like the proposed method, it produces the conditional distribution for the AoD and AoA.
\item Dedicated single beam training (DBT-single): The single beam transmission (with periodicity $T_d$) is used in combination with the proposed greedy recovery algorithm. This algorithm is included to verify the effectiveness of the dual beam transmission. 
\item Proposed dedicated dual beam training
\end{enumerate}
Fig. \ref{fig:overhead_new} provides the BER performance as a function of the training overhead for the several beam training schemes. By changing the periods of the beam training, $T_c$ and $T_d$, we change the training overhead. We set $\beta= 0.001$ and $0.002$ in Fig. \ref{fig:overhead_new} (a) and  (b), respectively. The rest of the system parameters are the same as those used  in Fig. \ref{fig:overhead_conv1}. We observe that the proposed scheme outperforms the competing schemes with the same training overhead.
Though both the proposed  and conventional beam cycling schemes can achieve the BER level being smaller than $10^{-3}$, the proposed scheme requires significantly less training overhead due to the short beam transmission period for the dedicated beams. 
In particular, when $T_d = 500, 200$ and $100$, the training overheads of the proposed method are $3.4$\%, $4.1$\%, and $5$\%, respectively. This is in contrast to the conventional beam cycling requiring $6.4$\%, $16$\%, and $32$\%  overhead for the same set of training periods $T_c = 500, 200$ and $100$.
Note that the training overhead of the proposed scheme can be reduced further when the beam transmission for the proposed scheme is performed only over the bandwidth occupied by the dedicated user. 
\begin{figure} [!t]
\centering 
	\includegraphics[width=3.8in,height=3.2in]{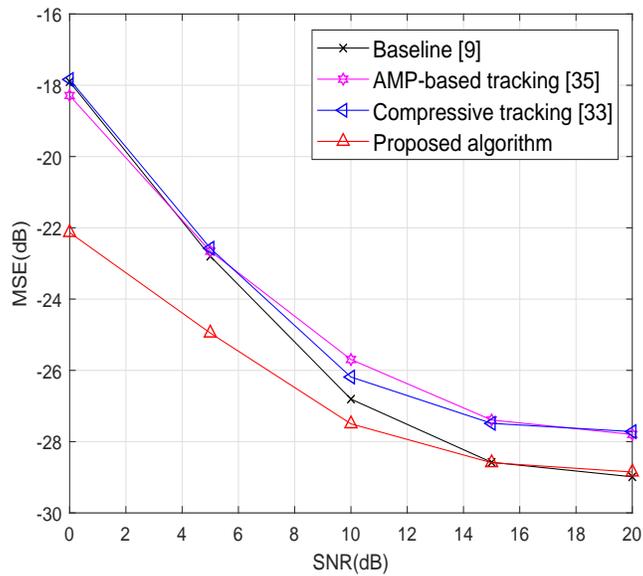}
\caption{Channel estimation MSE versus SNR } \label{fig:mse_performance2}
\end{figure}
Next, we compare the proposed scheme with the single beam transmission approach. Though the single beam transmission scheme has lower training overhead,  the proposed channel estimator (not to mention the OMP) does not work at all, as shown in Fig. \ref{fig:overhead_new}. 
We observe that the system matrix becomes rank-deficient with single beam transmission and thus the support cannot be  estimated accurately. The Kalman-based beam tracking scheme \cite{heathtr}  performs slightly better than OMP but it performs much worse than  the proposed method. Note that the proposed scheme also outperforms the POMDP-based beam design scheme. 

We evaluate the performance of the proposed channel estimation algorithm. Our algorithm is compared with the OMP algorithm \cite{chan_est}, the AMP-based channel tracking \cite{madhow}, and the compressive channel tracking \cite{sounding}. We use the OMP  algorithm as a baseline since our channel estimation algorithm reduces to the OMP when there exist no temporal correlation in AoD and AoA. Since our dedicated beam training optimizes the beams based on the probabilistic AoD and AoA information obtained by the channel estimator, other channel estimation methods are not compatible with the dedicated beam protocol. In order to compare the channel estimation performance only, we evaluate the algorithms  with common beam cycling with $T_c=100$ and $N_c=32$. 
We set the mobility parameter $\beta$ to $0.002$. 
The normalized mean square error (MSE) for the channel estimation is used as a performance metric. 
Fig. \ref{fig:mse_performance2} shows the MSE performance as a function of SNR. Note that the floor in the MSE for high SNR range is due to variation of the channels over the beam training period as well as the effect of quantizing the continuous AoD and AoA through virtual channel model. 
The proposed channel estimator outperforms the OMP  over the whole SNR range of interest. 
Note that this gain comes from the fact that the proposed algorithm exploits the temporal channel correlation across multiple beam training periods while the baseline algorithm independently processes for each beam training period. Note that such benefit of processing multiple measurement vectors would be greater in the presence of high noise level. As a result, we see large gap in performance between two methods as the SNR decreases.  Note that that the proposed algorithm also outperforms other state of the art beam tracking algorithms \cite{madhow,sounding}. 



\section{Conclusions}

In this paper, we have presented the novel beam training protocol that can significantly reduce training overhead for the mobility scenario in mmWave communications.
The proposed dedicated beam training is designed to perform beam training only for high mobility users.
Using a small number of training beams, the proposed dedicated beam training yields small training overhead. 
The proposed method employs the optimal beam selection strategy to find the best beams ensuring strong channel estimation performance for the target user.  In addition, we have devised the greedy sparse channel estimation algorithm that can exploits temporal channel correlation of the target user. 
Our numerical evaluations demonstrated that the proposed scheme achieves satisfactory system performance with small amount of training overhead as compared to the conventional methods.

\section{Appendix}
\subsection{Derivation of CRLB}
\label{subsec:app1}
In this subsection, we derive the CRLB for $\theta_{p,t,1}^{(b)}$.
   Since the channel parameters $\alpha_{p,t,1}$ and $\theta^{(b)}_{p,t,1}$ are unknown,  we use the information inequality for the complex variable $\alpha_{p,t,1}$ and the real variable $\theta^{(b)}_{p,t,1}$.  If we let $\underline{\theta} = [\alpha_{p,t,1}, \alpha_{p,t,1}^{*}, \theta^{(b)}_{p,t,1} ]$,  the CRLB of the real variable $\theta^{(b)}_{p,t,1}$ is given by \cite{ref:complex-CRLB}
\begin{align}
  CRLB &= ( \mathbf{V}^{-1})_{3,3} \\
  & =  [(\mathbf{V})_{3,3}-2Re\{(\mathbf{V})_{3,1}((\mathbf{V})_{1,1})^{-1}(\mathbf{V})_{1,3}\}]^{-1} \label{eq:vv}
\end{align}
where $\mathbf{V}$ is Fisher information matrix. The $(i,j)$th element of $\mathbf{V}$, $(\mathbf{V})_{i,j}$ is given by
\begin{align}
(\mathbf{V})_{i,j} 
=  -E\left[\frac{\partial}{\partial (\underline{\theta})_i^*}    \frac{\partial \log p(\mathbf{y}_{p,t}|\underline{\theta})}{\partial (\underline{\theta})_j} \right].
\end{align}
First, $(\mathbf{V})_{3,3}$ is obtained as
\begin{align}
(\mathbf{V})_{3,3} &= -E\left[\frac{\partial^2 \log p(\mathbf{y}_{p,t}|\underline{\theta})}{\partial \left( \theta_{p,t,1}^{(b)}\right)^{2}}\right]   \\
&= \frac{2}{\sigma_n^{2}}\left\| \alpha_{p,t,1} \mathbf{F}_{t}^{T}
\frac{{\partial}(\mathbf{a}^{(b)}(\theta_{p,t,1}^{b})^{*})}{\partial \theta_{p,t,1}^{b}}\right\|^{2}. \label{eq:I33 2}
\end{align}
The log-likelihood function $\log p(\mathbf{r}_{p,t}|\underline{\theta})$ can be expressed as
We can obtain the expression of $(\mathbf{V})_{1,3}$ and $(\mathbf{V})_{3,1}$ 
\begin{align}
(\mathbf{V})_{3,1} = (\mathbf{V})_{1,3}^{*} &= -E\left[\frac{\partial}{\partial \theta_{p,t,1}^{(b)}} \left( \frac{\partial \log p(\mathbf{y}_{p,t}|\underline{\theta})}{(\partial \alpha_{p,t,1}^{(b)})^*}\right)\right].  \\
&=\frac{1}{\sigma^{2}}\alpha_{p,t,1}  \frac{\partial\mathbf{a}^{(b)}(\theta_{p,t,1}^{(b)})}{\partial \theta_{p,t,1}^{(b)}} \mathbf{F}_{t}^{*} \mathbf{F}_{t}^{T} \mathbf{a}^{(b)}(\theta_{p,t,1}^{(b)})^{*}. \label{eq:I13 2}
\end{align}
Finally, $(\mathbf{V})_{1,1}$ can be obtained as
\begin{align}
(\mathbf{V})_{1,1} &= -E\left[\frac{\partial}{\partial \alpha_{p,t,1}} \left( \frac{\partial \log p(\mathbf{y}_{p,t}|\underline{\theta})}{(\partial \alpha_{p,t,1})^*}\right)\right]  \\
&= \frac{1}{\sigma^{2}}\left\| \mathbf{F}_{t}^{T} \mathbf{a}^{(b)}(\theta_{p,t,1}^{(b)})^{*} \right\|^{2}. \label{eq:I11 2}
\end{align}
Plugging (\ref{eq:I33 2}), (\ref{eq:I13 2}), and (\ref{eq:I11 2}) into (\ref{eq:vv}), we finally obtain the CRLB in (\ref{eq:CRLB}).
\end{document}